\documentclass[10pt,a4paper]{article}
\usepackage{t1enc}
\usepackage[latin1]{}			
\usepackage[english]{babel}
\begin{document}
\begin{center}
{\bf \huge
TOPOLOGY AND SOCIAL BEHAVIOUR OF AGENTS
}
\end{center}

\begin{center}
Ondrej Hudak, \footnote{\small present e-mail: hudako@mail.pvt.sk}
\end{center}

\begin{center}
Faculty of Finance, Matej Bel University, // Cesta na amfiteater 1, Banska Bystrica, Slovakia,//
 e-mail: hudak@financ.umb.sk
\end{center}

\begin{center}
Jana Tothova
\end{center}
\begin{center}
Stierova 23, SK-040 23 Kosice, Slovakia.
\end{center}
\newpage

\section{Abstract}

We call in a social group its members as agents. 
On a capital market the social group is formed from those who 
are buying and selling shares. In general the social group we 
characterise in accordance with 
T. Plummer / The Psychology of Technical Analysis, rev.ed., 
Probus Pub.Comp., Chicago, 1993/. We assume that individual behaviour 
of agents is influenced to some degree by the need to associate 
with other agents and to obtain the approval of other agents in 
the group. The group is characterised by a very large nonrational 
and emotional element to decisions of agents. It is due to the fact
 that making decisions an individual equates own needs with those
 of the other agents from the group. Any two agents from the group 
may interact. The interaction consists of the exchange of information 
and it costs some energy. The information is well defined, and we 
assume that agents interact in such a way that they give reference 
to the origin of the information if asked by other agents. Thus 
the agent may verify obtained  information. It is natural then 
that there exits a subgroup of interacting agents the interaction 
of which has the following property: it is nonreducible in the sense 
that in this subgroup there exists the interaction of a given agent 
with another one, this another one agent again interacts with another
 agent, and because it has no sense to exchange the information with 
the first one to verify the information, this last mentioned agent is 
different from the first one. This third agent can thus always verify 
the information from two sources either interacting with the first one,
 either he/she is interacting with the fourth agent. In the first case
 we have a minimal subgroup of interacting agents, which form a closed 
subroup in which every agent verifies information from two
sources - agents. In the second case the fourth agent either interacts
 with the first one and verifies exchanged information, either with 
the second one and verifies information. He of course can interact 
with the third one agent, however to verifiy information he needs 
again to interact with different agent from that interacting with 
which he obtained the new information. This is the reason why he 
interacts either with the first one agent, either with the second one. 
In the first case we have a closed subgroup of four agents in which 
every agent interacts with two agents. In the second case the closed
 subgroup reduces to the group of  three interacting agents /last 
three agents /, verify information interacting with the fifth agent. 
Then by the same procedure as in previous case we may obtain 
nonreducibile subgroups of three, of four and of five agents, or 
the fifth agent interacts with the sixth one. And so on. Thus we 
obtain closed subgroups of interacting agents in which every agent
 interacts with two other agents, and in which  the process of 
verification of information leads to closed linear structure. If any
 of agents from such an subgroup exchange and thus verifies information
 with any third one agent from the subroup, the structure of 
the subgroup changes, the subgroup becomes reducibile to two
 new nonreducibile subgroups. This is the process of differentiation 
of the first nonreducibile subgroup to two new ones. This is an 
example of an elementary transformation between two configurations 
of nonreducibile subroups. An vice versa: if the interaction between 
two agents which interact with three agents in the configuration of 
two nonreducibile subgroups vanishes, a new one nonreducibile subgroup
 creates and information is still verifiable. When the configuration 
in which an agent was interacting with three agents which were not
 interacting between themselves transform to a configuration in which 
two of the mentioned three agents interact, then we may observe 
the process of mitosis: a new nonreducibile group appears.Thus 
the transformation between this two configurations is reversibile. 
A cell is such a configuration of a given number of nonreducible 
subgroups in which every two interacting agents belongs to two 
nonreducibile subgroups /subgroups are connected in this sense
/ and which is closed . Such a cell  may disappear and may be created,
 may change number of nonreducible subroups in a reversible way.
 Becasue the structure, configuration of interactions between  
agents in the group, forms a macroscopic structure, we say that it is 
a microreversible process any process within a nonreducibile subgroup 
and within a cell. Statistical equilibrium of the whole group 
is characterized by a set of different subroups of the type mentioned
 above, and by a probability that such a subgroup occurs. Thus we have
 probability distribution which characterizes the group. Moreover 
there exists an equation of state which enables to compare different
 macroscopic states of the group.The statistical equilibrium due
 to microreversibility is characterised by the maximum of entropy 
and by the minumum of energy /costs of information exchange/. 
There exist constrains, such as a fixed number V of agents in 
the group,  a number E of interactions within the group, a number F 
of subgroups which are nonreducibile, and a number C of cells. Thus 
we have a structure which is equivalent to random cellular networks. 
Such networks and their evolution were described by
 N. Rivier /Physica 23D (1986) p. 129/. He applied methods 
of statistical mechanics to study these structures. We will use
 methods described by Rivier to study  social behaviour of agents, 
mainly the presence of topological structure of interactions between 
agents and its changes, which is the most important property of
 the group of agents. The area of a nonreducibile group which belongs 
to those nonreducibile groups which form the cell may be formed again 
for example by a sum of areas of agents characteristic areas. Note 
that area of the nonreducibile group may be also some other 
characteristics of the group of agents depending on studied social 
relations between agents. Thus we are able to study topology 
properties of interactions of agents. Their social behaviour is
 discussed. It can be shown that the equilibrium number of agents 
with which a given agent interacts is three for a group without 
cellls /the group forms a single cell/.  
\newpage

\tableofcontents{}

\newpage

\newpage
\section{Introduction}

We call in a social group its members as agents. 
In general the social group is characterised in accordance with 
T. Plummer \cite{1} . We assume that individual behaviour 
of agents is influenced to some degree by the need to associate 
with other agents and to obtain the approval of other agents in 
the group. The group is characterised by a very large nonrational 
and emotional element to decisions of agents. It is due to the fact
 that making decisions an individual equates own needs with those
 of the other agents from the group. Any two agents from the group 
may interact. The interaction consists of the exchange of information 
and it costs some energy. The information is well defined, and we 
assume that agents interact in such a way that they give reference 
to the origin of the information if asked by other agents. Thus 
the agent may verify obtained  information. It is natural then 
that there exits a subroup of interacting agents the interaction 
of which has the following property: it is nonreducible in the sense 
that in this subgroup there exists the interaction of a given agent 
with another one, this another one agent again interacts with another
 agent, and because it has no sense to exchange the information with 
the first one to verify the information, this last mentioned agent is 
different from the first one. This third agent can thus always verify 
the information from two sources either interacting with the first one,
 either he/she is interacting with the fourth agent. In the first case
 we have a minimal subgroup of interacting agents, which form a closed 
subroup in which every agent verifies information from two
sources - agents. In the second case the fourth agent either interacts
 with the first one and verifies exchanged information, either with 
the second one and verifies information. He of course can interact 
with the third one agent, however to verifiy information he needs 
again to interact with different agent from that interacting with 
which he obtained the new information. This is the reason why he 
interacts either with the first one agent, either with the second one. 
In the first case we have a closed subgroup of four agents in which 
every agent interacts with two agents. In the second case the closed
 subgroup reduces to the group of  three interacting agents /last 
three agents /. This process can continue futher, because the fourth 
agent may also verify information interacting with the fifth agent. 
Then by the same procedure as in previous case we may obtain 
nonreducibile subgroups of three, of four and of five agents, or 
the fifth agent interacts with the sixth one. And so on. Thus we 
obtain closed subroups of interacting agents in which every agent
 interacts with two other agents, and in which  the process of 
verification of information leads to closed linear structure. If any
 of agents from such an subgroup exchange and thus verifies information
 with any third one agent from the subroup, the structure of 
the subgroup changes, the subgroup becomes reducibile to two
 new nonreducibile subgroups. This is the process of differentiation 
of the first nonreducibile subgroup to two new ones. This is an 
example of an elementary transformation between two configurations 
of nonreducibile subroups. An vice versa: if the interaction between 
two agents which interact with three agents in the configuration of 
two nonreducibile subgroups vanishes, a new one nonreducibile subgroup
 creates and information is still verifiable. When the configuration 
in which an agent was interacting with three agents which were not
 interacting between themselves transform to a configuration in which 
two of the mentioned three agents interact, then we may observe 
the process of mitosis: a new nonreducibile group appears.Thus 
the transformation between this two configurations is reversibile. 
A cell is such a configuration of a given number of nonreducible 
subroups in which every two interacting agents belongs to two 
nonreducibile subgroups /subgroups are connected in this sense
/ and which is closed . Such a cell  may disappear and may be created,
 may change number of nonreducible subroups in a reversible way.
 Becasue the structure, configuration of interactions between  
agents in the group, forms a macroscopic structure, we say that it is 
a microreversible process any process within a nonreducibile subgroup 
and within a cell. Statistical equilibrium of the whole group 
is characterized by a set of different subroups of the type mentioned
 above, and by a probability that such a subroup occurs. Thus we have
 probability distribution which characterizes the group. Moreover 
there exists an equation of state which enables to compare different
 macroscopic states of the group.The statistical equilibrium due
 to microreversibility is characterised by the maximum of entropy 
and by the minumum of energy /costs of information exchange/. 
We will use
 methods of statistical physics to study  social behaviour of agents, 
mainly the presence of topological structure of interactions between 
agents and its changes, which is the most important property of
 the group of agents. There are three empiricaly observed dependencies of personal radius which enabled us to characterize the quantities of cells, faces, vertices and bonds \cite{2}.
There exist constrains, such as a fixed number V of agents in 
the group,  a number E of interactions within the group, a number F 
of subgroups which are nonreducibile, and a number C of cells. Thus 
we have a structure which is equivalent to random cellular networks. 
Such networks and their evolution were described by
 \cite{3}, \cite{4} N. Rivier /Physica 23D (1986) p. 129/. He applied methods 
of statistical mechanics to study these structures. We will use
 methods described by Rivier to study  social behaviour of agents, 
mainly the presence of topological structure of interactions between 
agents and its changes, which is the most important property of
 the group of agents. The area of a nonreducibile group which belongs 
to those nonreducibile groups which form the cell may be formed again 
for example by a sum of areas of  agents characteristic areas. Note 
that area of the nonreducibile group may be also some other 
characteristics of the group of agents depending on studied social 
relations between agents. Thus we are able to study topology 
properties of interactions of agents. Their social behaviour is
 discussed in. It can be shown that the equilibrium number of agents 
with which a given agent interacts is three for a group without 
cellls /the group forms a single cell/.

\section{Human groupings have hierarchical structure.} 

Human groupings have hierarchical structure, for an introduction see
\cite{1}. The civilisation consists of societes, societes consist of groups, and groups consist of individuals. A social group has power to rganise individuals and se them for its own purposes.
An agent we call a member of a social group, we assume that his/her special properties are defined.

Agent - its behaviour is influenced to some degree by the need to associate with other agents and to obtain the approval of other agents
in the group \cite{1}.
A group is chararcterized by a very large nonrational and emotional
elements in decisions of agents, and agent equates his/her 
needs with those of the other agents from the group [1]. 
Equates here means interacts, interaction costs some energy and consists of the exchange of information about his/her needs, 
the information is well defined.
Specification of agents here: agents interact in such a way that they
give reference to the origin of the information if asked by the other agent. Thus an agent may verify obtained information.

Nonreducible subgroup of the group of agents is a group which defined in the following way.
There exists interaction of a given agent with another one. This
another one agent interacts with another one. it has no sense for him/her to exchange information with the first one to verify
information /to obtain approval of at least another one agent/:
thus the agent interacts with and agent diferent from the first one.
This the third agent can verify the information from two sources: 
either interacting with the first one, either interacting with the fourth one.
The first case leads to formation of a nonreducibile subgroup of three 
agents in which every agent verifies information from two different 
sources /we asuem that information is verified if an agent verifies
information from at least two different sources/.
The second case: the fourth agent interacts either with the first
one, and we ontain a nonreducibile subgroup of four agents, either this agent interacts with the second agent,
and we obtain a nonreducibile subgroup of three agents, this process
may continue further: the fourth agent may verifiy information
interacting with the first ones, the process just describes now may
lead to a nonreducibile subgroup with three, four or five agents,
etc. Thus in the group of interacting agents there exist nonreducibile
subgroups of agents which are closed as concerning exchange of an information and in which every agent interacts with two
and only two other agents, thus the structure of interaction of agents is "linear-circular".

\section{Nonreducibile subgroups of the group of agents may transform.}

Reduction of a nonreducibile subgroup to two nonreducibile subgroups:
if any agent from a given nonreducibile subgroup interacts as
 concerning exchange of an information 
with a third one agent from the nonreducibile subgroup.
And vice versa: if the interaction as concerning exchange of an information between two agents in the configuration of two nonreducibile subgroups vanishes, a new one nonreducibile subgroup appears. Mitosis is a process in which an agent which was interacting with three agents which were not interacting between themselves transforms to a configuration in which two of the mentioned three agents interact, then a new nonredcibile subgroup appears, this process is reversibile.

Cell is a configuration of a given number of nonreducibile subgroups
in which everz two interacting agents belong to two nonredcibile
subgroups of a closed subgroup formed from nonreducibile subgroups.
Cella of the group of agents may reversible transform, they may 
disappear, may be created, may change a number of nonreducibile
subgroups.

\section{The structure of the group is macroscopic}

The group has a given number of agents, of nonreducibile subgroups and of cells as concerning exchange of information. Statistical 
equilibrium of the group is characterized by different macroscopic 
structure as concerning exchange of an information and by a probability that a nonreducible group appears, this probability characterizes the group. Equation of state comparison of different 
macroscopic states of the group. Statistical equilibrium exist in the group due to reversibility, it is characterized by the maximum of entropy and by the minimum of energy /costs of information exchange/.
Constraints are given by a fixed number V of agents in the group, a number E of interactions, a number F of subgroups which are nonreducibile, a number C of cells, note that conservation law holds \cite{4}:
\begin{equation}
\label{1}
- C + F - E + V = 1
\end{equation}

Note that structural stability exists \cite{4}: only agents with interaction with four agents are structuraly stable in 3d, and only agents with interaction with three agents are structuraly stable in 2d. This structure is equivalent to random cellular networks,
such networks and their evolution were described by N. RIVIER \cite{3}, \cite{4} by methods statistical mechanics.

We will use these methods to study social behaviour of agents,
and changes of this behaviour, mainly the presence of the topological
structure of interactions between agents and its changes, the most 
important property of the group of agents which coresponds to maximum 
of informational entropy.

One can define an area of nonreducibile subgroup and volume of the cell. We asume that there is homogeneity, and no costs of information.

\section{Personal area and social behaviour of agents of the group}

The average area A(n) of an n-sided cell is \cite{3}, \cite{4} / if A is the total area in which group is localised/:
\begin{equation}
\label{2}
A(n) = \frac{A}{F}\lambda (n-(6-\frac{1}{\lambda}))
\end{equation}

Asume that this area corresponds to a personal area of agents forming 
an n-sised nonreducibile subgroup, due to homogeneity and equilibrium
every agent contributes $\frac{1}{3}$ personal area A(n), thus $ \lambda = \frac{1}{6}$ and
\begin{equation}
\label{3}
A(n) = \frac{A}{F} \frac{n}{6}
\end{equation}

One agent contributes $ \frac{1}{3} $ of his personal area to the area
A(n), if we denote r the radius of the agent's personal area then:
\begin{equation}
\label{4}
\frac{\pi r^{2}}{3} = \frac{A}{6F}
\end{equation}

As we see personal radius increases as a square root with increased

total area A per a nonreducibile group. As we see personal radius
decreases as an inverse square root with increasing number of
nonreducibile groups, in towns personal area is observed to be smaller
than in villages \cite{2}, this corresponds to smaller $ \frac{A}{F} $
ratio in towns than in villages according to our equation above,
the ratio $ \frac{A}{F} $ is more-less constant in area A in these cases, however it is increasing with the density of agents in the group (town, village) and thus number of nonreducibile groups is increasing in our equation for personal area qualitatively, personal radius is increasing with decreasing risk which a person expects
\cite{2}: intimous - smaller radius is 0.15 m and less for intimous contacts /love, security, ... /, intimous - larger radius is 0.15 m to 0.45 m for less intimous contacts /relatives/, personal - smaller radius is 0.45 m to 0.90 m for personal contacts with close friends and relatives, personal - larger radius is 0.90 m to 1.20 m for personal contacts with friend, bussines people, neoghbours.
Thus we see from our equation that F may be associated with risk:
for constant area A the larger F the smaller r and smaller acceptable risk (*). Personal radius is large for inhabitants \cite{2} of New Zeeland, Australia and white North-Americans, it is middle for inhabitants of Great Britain, Switzerland, Sverige, Germany, Austria
and it is small for inhabitants of Arab countries, Japans, South-Americans, inhabitants of countries around Middleterrenien sea /Italy, France, Greece, ... / and black North-Americans: one can say qualitatively that (probably due to temperament) the first group is characterised by low risk activities and by their preference, the second group is characterised by middle risk activities, and the last group is characterised by high risk activities. Thus first group has F lower than the second group, and second group lower than the third group: this is consistent with the statement (*).

\section{Conclusion}

We conclude that the three mentioned empiricaly observed dependencies
of personal radius dependence on some factors enabled us to characterize the quantity F as the quantity which characteriyes verification of information, the smaller personal radius the larger process of verification of verbal and non-verbal information about the other person in other to accept the risk that this person will be closer to us as a person. We expect that this conclusion is a general conclusion.

\section{Acknowledement}

The paper represents a part of results of the VEGA project 1/0495/03.

\end{document}